\documentclass{PoS}

\usepackage{latexsym} 
\usepackage{longtable} 
\usepackage{mathrsfs}
\usepackage[geometry]{ifsym}
\usepackage{epsfig}
\usepackage{pstricks,pst-grad}
\usepackage{amsmath,amsfonts,amssymb}
\usepackage{graphicx,rotating,psfrag}
\usepackage{comment}

\newcommand{\mb}[1]{\mathbf{#1}}
\newcommand{\mc}[1]{\mathcal{#1}}
\newcommand{\mr}[1]{\mathrm{#1}}
\newcommand{\bra}[1]{\langle#1|}
\newcommand{\ket}[1]{|#1\rangle}
\renewcommand{\mathcal}[1]{\mathscr{#1}}

\newrgbcolor{orange}{1.0 0.647059 0.0}
\newrgbcolor{newmagenta}{1.0 0.0 1.0}
\newrgbcolor{darkgreen}{0.0 0.545098 0.0}

\title{Charm semileptonic decays and $|V_{cs(d)}|$ from heavy clover quarks and 2+1 flavor asqtad staggered ensembles}

\ShortTitle{$D\to K(\pi)\ell\nu$ and $|V_{cs(d)}|$ from heavy clover on 2+1 flavor asqtad MILC ensembles}

\author{%
\speaker{Jon~A.~Bailey},$^{a}$
D.~Du,$^{b}$
A.X.~El-Khadra,$^b$
Steven~Gottlieb,$^c$
R.D.~Jain,$^b$
A.S.~Kronfeld,$^d$
R.S.~Van~de~Water,$^d$
and
R.~Zhou$^c$ \\
Email:  \email{jabsnu@gmail.com}\\ \\
\llap{$^a$}Department of Physics and Astronomy, Seoul National University, Seoul, 151-747, ROK\\
\llap{$^b$}Physics Department, University of Illinois, Urbana, IL  61801, USA\\
\llap{$^c$}Department of Physics, Indiana University, Bloomington, IN  47405, USA\\
\llap{$^d$}Theoretical Physics Department, Fermilab, Batavia, IL  60510, USA
}
\author{
Fermilab Lattice and MILC Collaborations}

\abstract{
By combining experimentally measured partial branching fractions for the semileptonic decays $D\to K\ell\nu$ and $D\to\pi\ell\nu$ with lattice calculations of the form factors $f_+^{D\to K}(q^2)$ and $f_+^{D\to\pi}(q^2)$, one can extract the CKM matrix elements $|V_{cs}|$ and $|V_{cd}|$.
We are calculating the form factors by using Fermilab charm and asqtad staggered light and strange quarks on 2+1 flavor asqtad staggered ensembles generated by the MILC Collaboration.  
We vary the light valence quark masses from $0.4m_s$ to $0.05m_s$ ($m_s$ is the strange sea-quark mass), and the lattice spacings, from about $0.12\ \text{fm}$ to about $0.045\ \text{fm}$.  
We extrapolate to the physical light-quark mass and the continuum limit using heavy-light meson staggered chiral perturbation theory about the SU(2) and SU(3) limits, compare the resulting (preliminary) form factor shapes with experiment, and discuss our errors.
}

\FullConference{The 30th International Symposium on Lattice Field Theory\\
                 June 24 -- 29,  2012\\
                 Cairns, Australia}

\begin{document}

\section{Introduction}

Tests of second row and column unitarity are limited by the uncertainty in $|V_{cs}|$, determined from (semi)leptonic decays, while assuming three-generation unitarity leads to the very precise (Standard Model) values for $|V_{cs(d)}|$~\cite{PDG}.  Direct determinations of $|V_{cs(d)}|$ are being improved by combining increasingly precise lattice calculations of decay constants and form factors~\cite{FNALMILC,HPQCD} and information from experiment~\cite{BABAR,CLEO}.  The results have been used to improve tests of unitarity and, by comparing with the Standard Model values, to validate lattice calculations of $B$ decay constants and form factors.  {\it Ab initio} calculations of hadronic weak matrix elements of $B$ decays are central in a number of ongoing searches for new physics~\cite{NPBdecays}.

In the rest frame of the $D$ meson, the semileptonic branching fractions and CKM matrix elements are related by
\begin{equation}\label{eq:DSLdecay}
\frac{d\Gamma(D\to P\ell\nu)}{dq^2}=\frac{G_F^2}{24\pi^3}|V_{cx}|^2|\mathbf{p}_P|^3|f_+^{D\to P}(q^2)|^2,
\end{equation}
where we neglect the lepton masses, $x=s,d$ denotes the strange or down quark, $P=K,\pi$ is the daughter meson, $q^2\equiv(p_D-p_P)^2$ is the invariant mass of the leptons, and the form factor $f_+^{D\to P}(q^2)$ is defined in terms of the hadronic matrix element of the flavor-changing vector current $\mc{V}_\mu=i\bar{x}\gamma_\mu c$:
\begin{equation}
\langle P|\mc{V}|D\rangle=f_+(q^2)\left(p_D+p_P-\frac{m_D^2-m_P^2}{q^2}q\right) + f_0(q^2)\frac{m_D^2-m_P^2}{q^2}q.\label{eq:hadffcont}
\end{equation}
Given the partial branching fractions, $d\Gamma/dq^2$, from experiment and the form factor normalizations, {\it e.g.}, $f_+^{D\to P}(q^2=0)$, from theory, one can use Eq.~(\ref{eq:DSLdecay}) to extract the CKM matrix elements $|V_{cx}|$.

Below we briefly describe our method and data set before focusing on the current status of the chiral-continuum extrapolation, preliminary results for the form factor shapes, and our anticipated errors.  At present the analyses are blinded; that is, the absolute normalizations of the form factors are hidden.

\section{Method}

For lattice calculations a convenient parametrization of the form factors is
\begin{equation}
\bra{P}V\ket{D}=\sqrt{2m_D}\left(v f_{\|}(E_P)+p_{\bot}f_\bot(E_P)\right),\label{eq:ffdef}
\end{equation}
where $v=p_D/m_D$ is the 4-velocity of the $D$ meson, $p_\perp\equiv p_P-(p_P\cdot v)v$, $E_P=(m_D^2+m_P^2-q^2)/(2m_D)$ is the energy of the recoiling $P$ meson, and $V_\mu$ is the lattice version of the flavor-changing vector current $\mc{V}_\mu$.

We construct the lattice current out of light staggered and heavy clover fields~\cite{Wingate:2002fh}.
To match the continuum currents, the lattice current must be properly normalized.  The matching factors $Z_{V_{cx}^\mu}$ nearly equal the geometric mean of degenerate vector current renormalization factors $Z_{V_{xx}^4}$ and $Z_{V_{cc}^4}$, which we are calculating nonperturbatively~\cite{FNALMILC}.  The correction factors $\rho_{V_{cx}^\mu}$ are constructed to be close to unity, and a subset of our collaboration has calculated them in one-loop lattice perturbation theory~\cite{Aidaetal}:
\begin{equation}
\bra{P}\mc{V}^\mu\ket{D}=Z_{V_{cx}^\mu}\bra{P}V_{cx}^\mu \ket{D},\quad
Z_{V_{cx}^\mu}=\rho_{V_{cx}^\mu}\sqrt{Z_{V_{xx}^4}Z_{V_{cc}^4}},\quad
\rho_{V_{cx}^\mu} = 1+\mc{O}(\alpha_s).
\end{equation}
Although the deviation of $\rho_{V_{cx}^\mu}$ from one is formally $\mc{O}(\alpha_s)$, its numerical coefficient is small, and the $\mc{O}(\alpha_s)$ correction, very small.  
To protect against analyst bias, we blind the analyses by introducing multiplicative offsets in $\rho_{V_{cx}^\mu}$.

The matrix elements $\bra{P}V_{cx}^\mu\ket{D}$ can be obtained from 3-point and 2-point correlators of the $P$ and $D$ mesons.  We extract $\bra{P}V_{cx}^\mu\ket{D}$ from the correlator ratio
\begin{equation}
\frac{1}{\phi_{P\mu}}\frac{{\overline{C}}_{3,\mu}^{D\to P}(t,T;\mb{p}_P)}{\sqrt{{\overline{C}_2^P(t;\mb{0}}){\overline{C}}_2^D(T-t)}}\frac{E_P}{e^{-E_Pt}}\sqrt{\frac{2e^{-m_P t}}{e^{-m_D(T-t)}}},\label{eq:new_ratio}
\end{equation}
where $\phi_{P\mu}\equiv(1,\ \mb{p}_P)$, $C_2^P$ and $C_2^D$ are 2-point $P$- and $D$-meson correlators, respectively, and $C_3^{D\to P}$ is a 3-point correlator in which the flavor-changing vector current destroys the $D$ meson at rest and creates the $P$ meson with momentum $\mb{p}_P$.  $T$ and $t$ are the source-sink separation and current insertion time in $C_3$, respectively, and ${\overline C}_3,\ {\overline C}_2$ are averages of 3-point and 2-point correlators; the averages were designed to suppress oscillations from opposite-parity states~\cite{btopipaper}.
For insertion times $t$ far from source and sink, $1\ll t\ll T$, the temporal and spatial components of the ratio respectively approach the form factors $f_{\|}$ and $f_\bot$.

\section{Data}
We analyze approximately unitary data generated on MILC asqtad ensembles with several sea-quark masses $\geq 0.05m_s$ and four lattice spacings $\lesssim 0.12\ \mr{fm}$ (Table~\ref{tab:ens}).  
The $P$-meson 2-point correlators and 3-point correlators are generated at momenta $\mathbf{p}_P=(0,0,0)$, $(1,0,0)$, $(1,1,0)$, $(1,1,1)$, $(2,0,0)$, in units of $2\pi/aL$.  The $D$-meson 2-point correlators are generated with the $D$ meson at rest.
We use local interpolating operators for the $P$ 2-points, and smear the $D$ interpolating operators in the 2-point and 3-point correlators with a 1S charmonium wavefunction~\cite{btopipaper}.
To increase statistics, the correlators are generated at four source times spaced evenly along the time direction.  Autocorrelations are suppressed by randomizing the spatial location of the source with configuration.
We generate the 3-point correlators at four source-sink separations $T$ to construct the averages ${\overline C}_3$~(Eqs.~(37,38) of Ref.~\cite{btopipaper}) and to minimize errors while avoiding excited state contamination~\cite{lat2009proc}.
%
\begin{table}
\begin{center}
\begin{tabular}{cr@{$^3\times$}lr@{/}lcr|r@{/}lr}
    \hline\hline
    $\approx a\ \mathrm{(fm)}$ & \multicolumn{2}{c}{$L^3\times n_t$} & \multicolumn{2}{c}{$am_l^\mathrm{sea}/am_s^\mathrm{sea}$} & Key & $N_\mathrm{conf}$ & 
    \multicolumn{2}{c}{$am_l^\mathrm{val}/am_s^\mathrm{val}$} & $\kappa_c$ \\ 
    \hline 
    $0.12$ & 20&64 & 0.02&0.05 & {\blue \Square} & 2052 & 0.02&0.0349 & 0.1259 \\ 
           & 20&64 & 0.01&0.05 & {\green \Square} & 2259 & 0.01&0.0349 & 0.1254 \\ 
           & 20&64 & 0.007&0.05 & {\orange \Square} & 2110 & 0.007&0.0349 & 0.1254 \\ 
           & 24&64 & 0.005&0.05 & {\red \Square} & 2099 & 0.005&0.0349 & 0.1254 \\ 
    \hline 
    $0.09$ & 28&96 & 0.0124&0.031 & {\blue \Circle} & 1996 & 0.0124&0.0261 & 0.1277 \\ 
           & 28&96 & 0.0062&0.031 & {\green \Circle} & 1931 & 0.0062&0.0261 & 0.1276 \\ 
           & 32&96 & 0.00465&0.031 & {\orange \Circle} & 984 & 0.00465&0.0261 & 0.1275 \\ 
           & 40&96 & 0.0031&0.031 & {\red \Circle} & 1015 & 0.0031&0.0261 & 0.1275 \\ 
           & 64&96 & 0.00155&0.031 & {\newmagenta \Circle} & 791 & 0.00155&0.0261 & 0.1275 \\ 
    \hline 
    $0.06$ & 48&144 & 0.0072&0.018 & {\blue \TriangleUp} & 593 & 0.0072&0.0188 & 0.1295 \\ 
           & 48&144 & 0.0036&0.018 & {\green \TriangleUp} & 673 & 0.0036&0.0188 & 0.1296 \\ 
           & 56&144 & 0.0025&0.018 & {\orange \TriangleUp} & 801 & 0.0025&0.0188 & 0.1296 \\ 
           & 64&144 & 0.0018&0.018 & {\red \TriangleUp} & 827 & 0.0018&0.0188 & 0.1296 \\ 
    \hline
    $0.045$ & 64&192 & 0.0028&0.014 & {\darkgreen \TriangleDown} & 801 & 0.0028&0.0130 & 0.1310 \\ 
    \hline\hline
\end{tabular}
\end{center}
\caption{\label{tab:ens}Ensembles, light valence masses, and hopping parameters.  The up-down valence mass is equal to the up-down sea mass, while the strange valence mass is approximately physical, tuned via the $K$ mass.  The charm mass is also approximately physical, tuned via the $D_s$ mass.  ``Key'' is the legend for plots in Sec.~4.}
\end{table}
%

\section{\label{sec:extrap}Chiral-continuum extrapolation}
For sufficiently small energies, quark masses, and lattice spacings, S$\chi$PT describes the energy, quark-mass, and lattice-spacing dependence of the form factor data in a model-independent way.  Below we show fits of data through $\mb{p}_P=(1,1,0)$ to SU(3) and SU(2) S$\chi$PT.  The key to the plots is in Table~\ref{tab:ens}.
\begin{figure}[tbph]
\begin{minipage}[c]{0.47\textwidth}
\begin{center}
\psfrag{r1^(-1/2)*pion_fperp}[b][b][1.0]{$r_1^{-1/2}f_\bot^{D\to\pi}$}
\psfrag{r1*Epi}[][][1.0]{$r_1E_\pi$}
\psfrag{chisq}[][][0.9]{$\chi^2/28=1.15$, $p=0.270$}
\includegraphics[clip,width=1.0\textwidth]{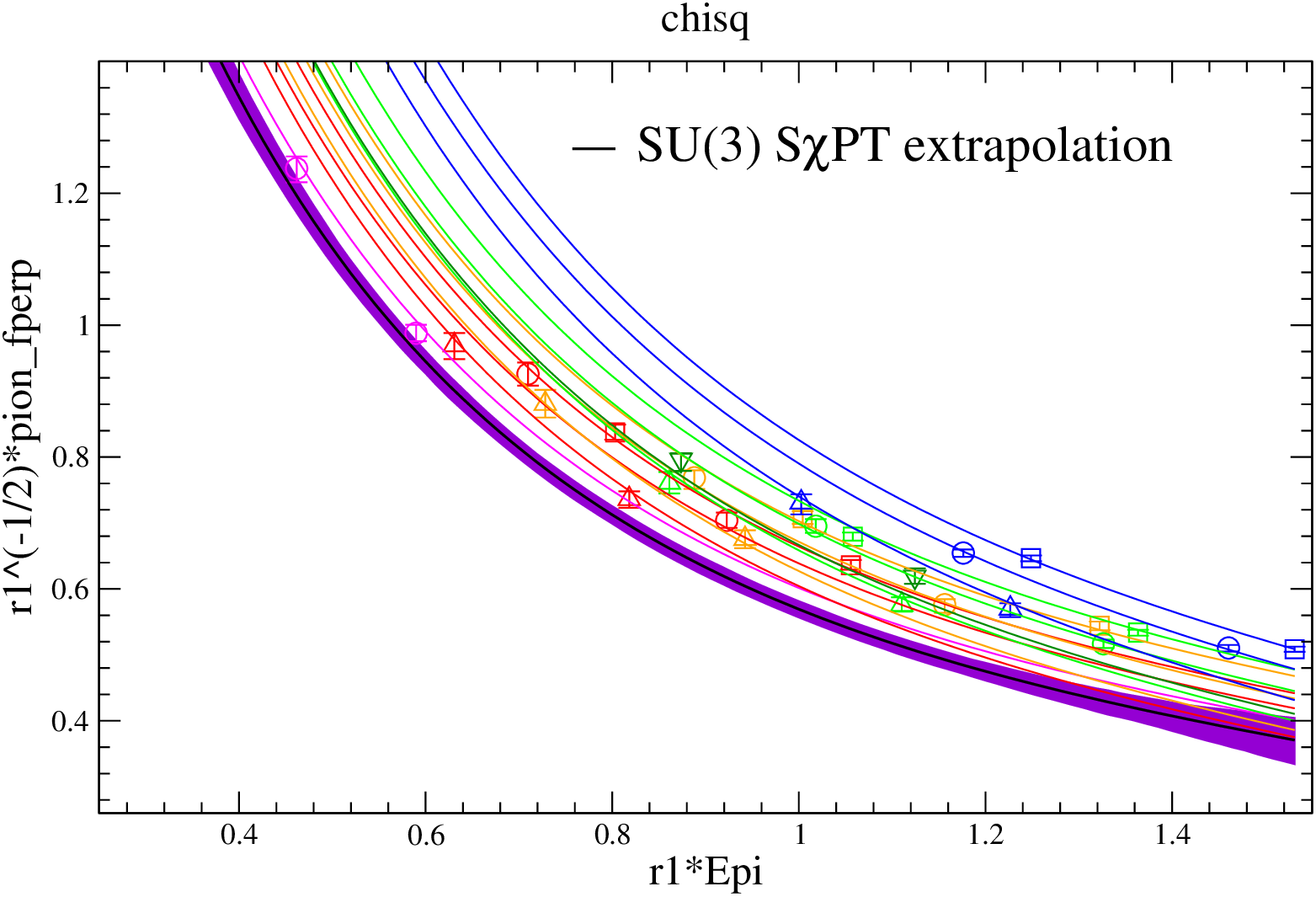}
\end{center}
\end{minipage}
\hspace{0.05\textwidth}
\begin{minipage}[c]{0.47\textwidth}
\begin{center}
\psfrag{r1^(-1/2)*pion_fperp}[b][b][1.0]{$r_1^{-1/2}f_\bot^{D\to\pi}$}
\psfrag{r1*Epi}[][][1.0]{$r_1E_\pi$}
\psfrag{chisq}[][][0.9]{$\chi^2/28=1.10$, $p=0.330$}
\includegraphics[clip,width=1.0\textwidth]{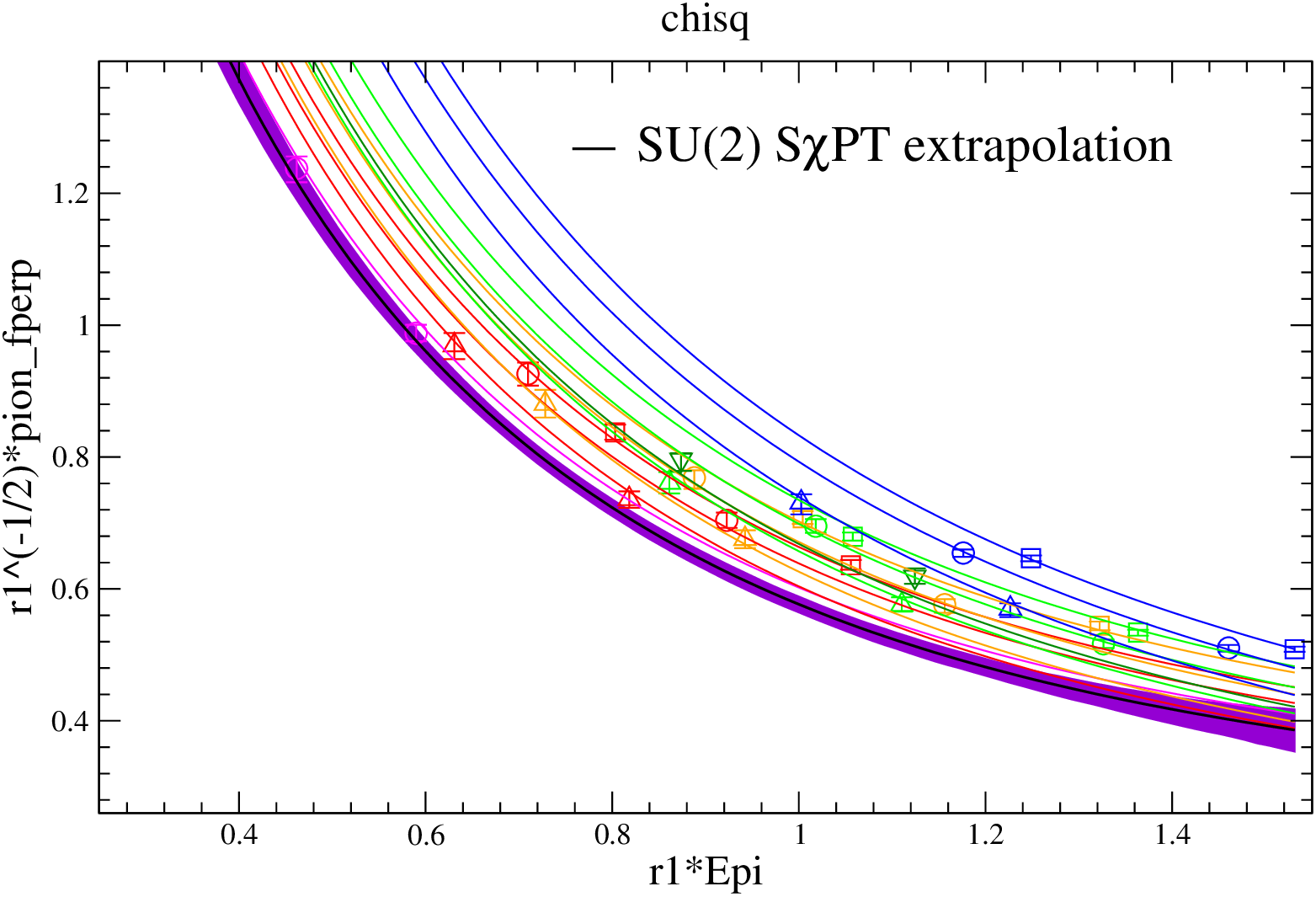}
\end{center}
\end{minipage}
\caption{\label{fig:pion_fperp}Fits of $f_\bot^{D\to\pi}$ data to SU(3) (left) and SU(2) (right) S$\chi$PT.  The fit functions include the chiral logarithms and analytic terms at NLO and analytic terms at NNLO.  Errors are statistical, from bootstrap ensembles.}
\end{figure}
\begin{figure}[tbph]
\begin{minipage}[c]{0.47\textwidth}
\begin{center}
\psfrag{r1^(1/2)*pion_fpar}[b][b][1.0]{$r_1^{1/2}f_{\|}^{D\to\pi}$}
\psfrag{r1*Epi}[][][1.0]{$r_1E_\pi$}
\psfrag{chisq}[][][0.9]{$\chi^2/42=1.53$, $p=0.0157$}
\includegraphics[clip,width=1.0\textwidth]{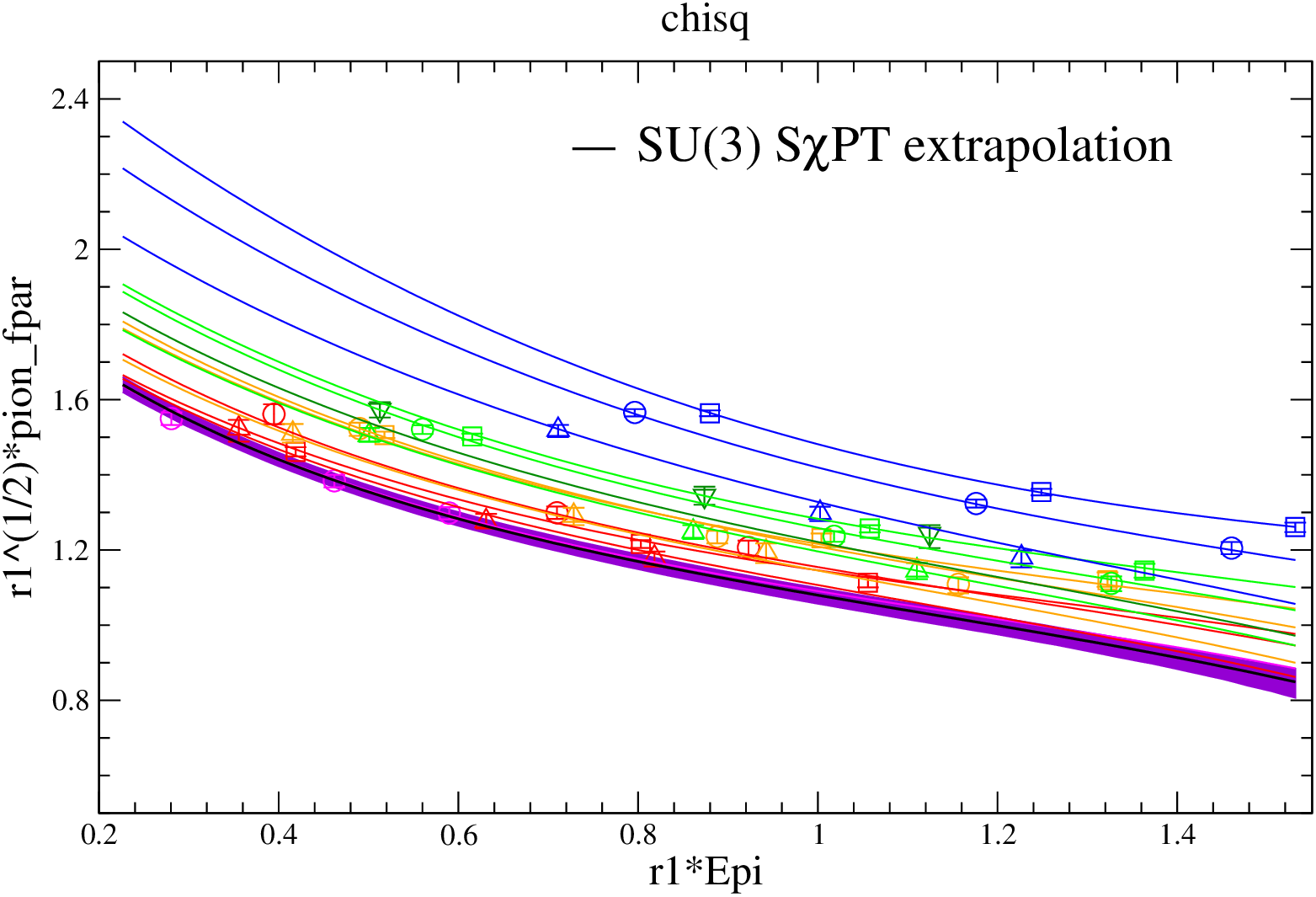}
\end{center}
\end{minipage}
\hspace{0.05\textwidth}
\begin{minipage}[c]{0.47\textwidth}
\begin{center}
\psfrag{r1^(1/2)*pion_fpar}[b][b][1.0]{$r_1^{1/2}f_{\|}^{D\to\pi}$}
\psfrag{r1*Epi}[][][1.0]{$r_1E_\pi$}
\psfrag{chisq}[][][0.9]{$\chi^2/42=1.36$, $p=0.0603$}
\includegraphics[clip,width=1.0\textwidth]{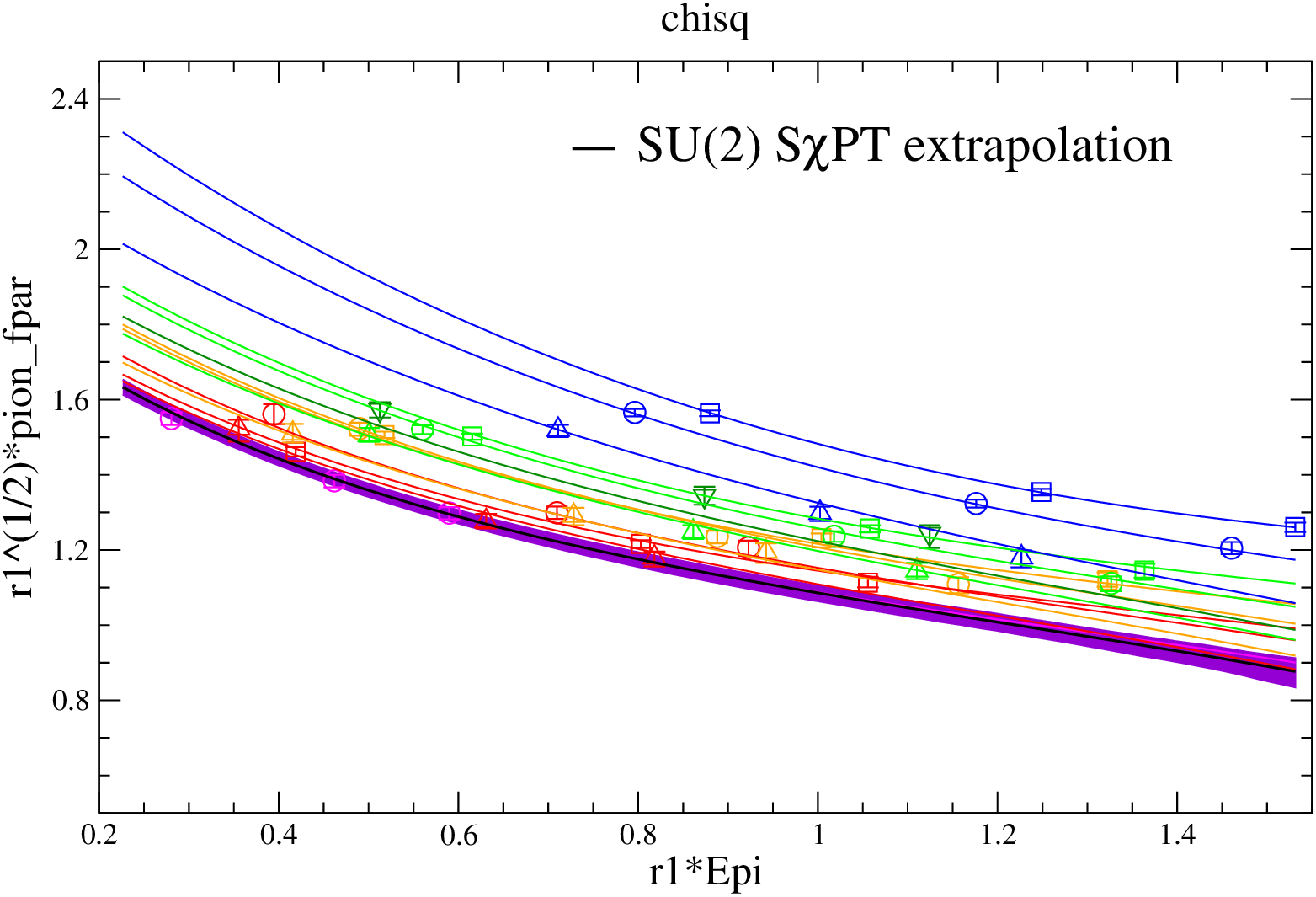}
\end{center}
\end{minipage}
\caption{\label{fig:pion_fpar}Fits of $f_{\|}^{D\to\pi}$ data to SU(3) (left) and SU(2) (right) S$\chi$PT.  The fit functions include the chiral logarithms and analytic terms at NLO and analytic terms at NNLO.  Errors are statistical, from bootstrap ensembles.}
\end{figure}
\begin{figure}[tbph]
\begin{minipage}[c]{0.47\textwidth}
\begin{center}
\psfrag{r1^(-1/2)*kaon_fperp}[b][b][1.0]{$r_1^{-1/2}f_\bot^{D\to K}$}
\psfrag{r1*EK}[][][1.0]{$r_1E_K$}
\psfrag{chisq}[][][0.9]{$\chi^2/28=0.858$, $p=0.680$}
\includegraphics[clip,width=1.0\textwidth]{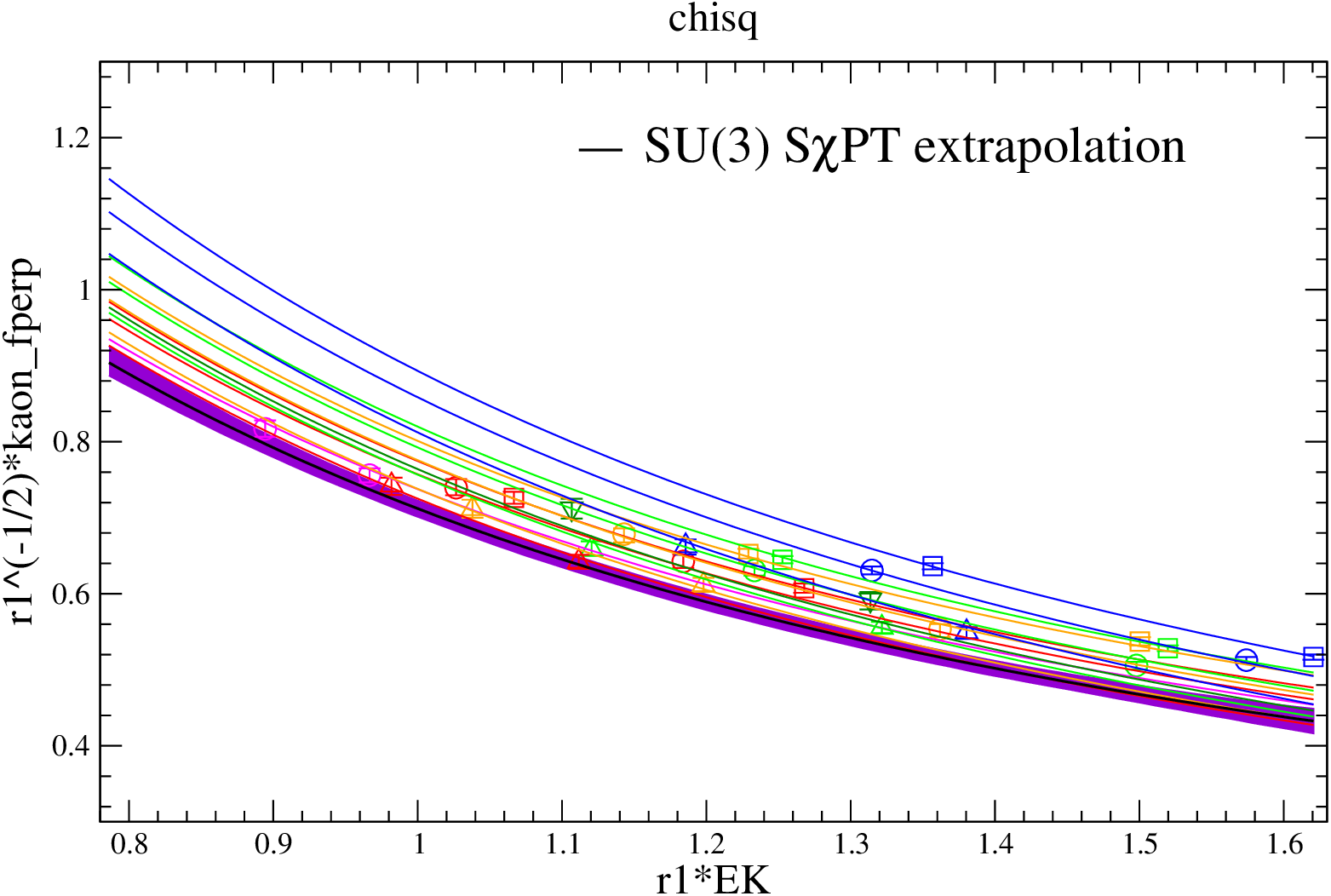}
\end{center}
\end{minipage}
\hspace{0.05\textwidth}
\begin{minipage}[c]{0.47\textwidth}
\begin{center}
\psfrag{r1^(-1/2)*kaon_fperp}[b][b][1.0]{$r_1^{-1/2}f_\bot^{D\to K}$}
\psfrag{r1*EK}[][][1.0]{$r_1E_K$}
\psfrag{chisq}[][][0.9]{$\chi^2/28=0.727$, $p=0.851$}
\includegraphics[clip,width=1.0\textwidth]{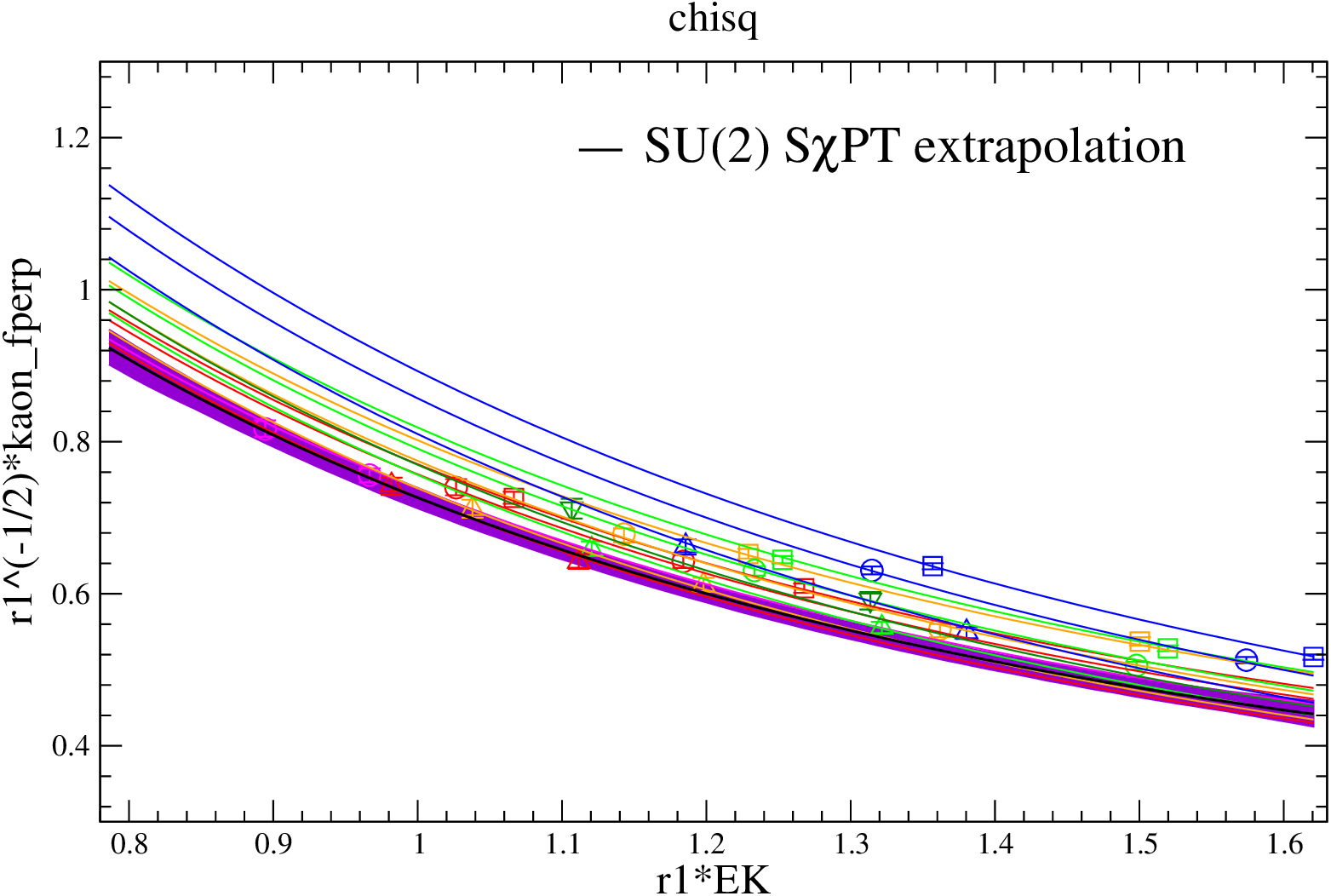}
\end{center}
\end{minipage}
\caption{\label{fig:kaon_fperp}Fits of $f_\bot^{D\to K}$ data to SU(3) (left) and SU(2) (right) S$\chi$PT.  The fit functions include the chiral logarithms and analytic terms at NLO and analytic terms at NNLO.  Errors are statistical, from bootstrap ensembles.}
\end{figure}
\begin{figure}[tbph]
\begin{minipage}[c]{0.47\textwidth}
\begin{center}
\psfrag{r1^(1/2)*kaon_fpar}[b][b][1.0]{$r_1^{1/2}f_{\|}^{D\to K}$}
\psfrag{r1*EK}[][][1.0]{$r_1E_K$}
\psfrag{chisq}[][][0.9]{$\chi^2/42=1.20$, $p=0.176$}
\includegraphics[clip,width=1.0\textwidth]{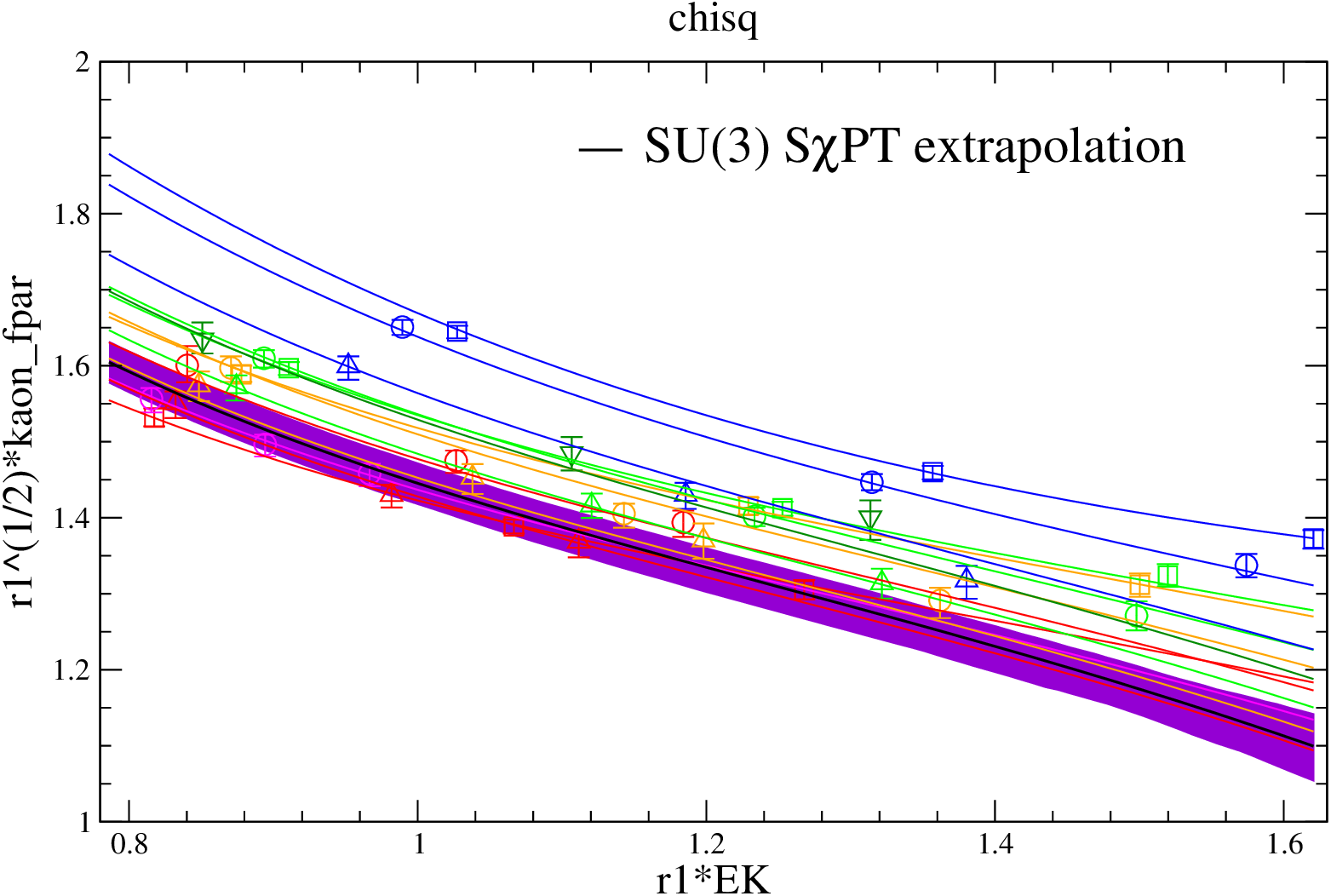}
\end{center}
\end{minipage}
\hspace{0.05\textwidth}
\begin{minipage}[c]{0.47\textwidth}
\begin{center}
\psfrag{r1^(1/2)*kaon_fpar}[b][b][1.0]{$r_1^{1/2}f_{\|}^{D\to K}$}
\psfrag{r1*EK}[][][1.0]{$r_1E_K$}
\psfrag{chisq}[][][0.9]{$\chi^2/42=0.866$, $p=0.716$}
\includegraphics[clip,width=1.0\textwidth]{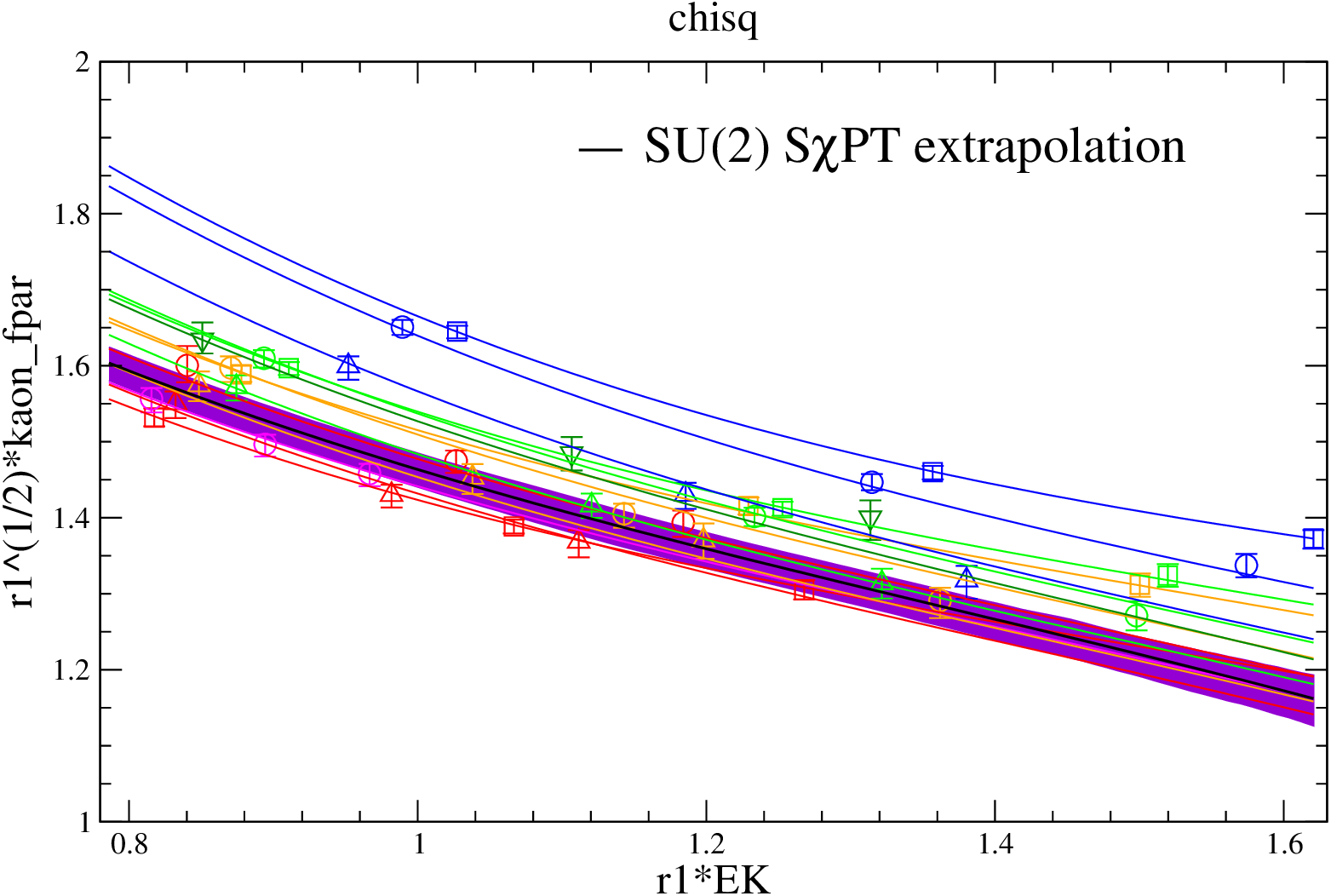}
\end{center}
\end{minipage}
\caption{\label{fig:kaon_fpar}Fits of $f_{\|}^{D\to K}$ data to SU(3) (left) and SU(2) (right) S$\chi$PT.  The fit functions include the chiral logarithms and analytic terms at NLO and analytic terms at NNLO.  Errors are statistical, from bootstrap ensembles.}
\end{figure}
The fits in Figs.~\ref{fig:pion_fperp}--\ref{fig:kaon_fpar} are fits to the data for $f_\bot$ and $f_{\|}$ separately; we have not imposed the kinematic constraint at $q^2=0$.  Restricting the data in the fit to have momenta less than $(1,1,1)$ ensures reasonable behavior in the chiral expansion even on the coarse $0.4m_s$ ensemble, where the corresponding recoil energies are greatest.

We obtain the SU(2) S$\chi$PT fit functions by integrating out the strange quark in SU(3) S$\chi$PT~\cite{AB_SHMChPT,Ran_note_sharpe}.
We include the strange-quark mass dependence in the SU(2) S$\chi$PT LECs by expanding in the deviation from the physical strange-quark mass.  The NLO terms absorb the scale dependence of the chiral logarithms, and we also include the analytic NNLO terms.

The SU(2) fits above have greater $p$-values than the SU(3) fits; the difference is pronounced for $f_{\|}^{D\to K}$ (Fig.~\ref{fig:kaon_fpar}).  However, the fit results are not obviously superior, and in some cases the chiral-continuum extrapolated values differ by a few percent.

The fits to $f_{\|}^{D\to\pi}$ have somewhat small $p$-values.
At higher energies, the errors of the chiral-continuum extrapolated curves for $f_\bot^{D\to\pi}$ grow, and the curves for $f_{\|}^{D\to\pi}$ and $f_{\|}^{D\to K}$ exhibit apparently unphysical inflection.  These behaviors at higher energies may reflect the absence of data from the more chiral and finer ensembles; adding data at higher momenta from these ensembles may decrease the errors and eliminate this inflection.

The factors $Z_{V_{cc}^4}$ and $Z_{V_{xx}^4}$ are preliminary, and we are updating them.  Before the final fits, the data must be shifted to the retuned $\kappa_c$ values.  We have generated additional data on one of the coarse ensembles to correct for the error in $\kappa_c$-tuning and estimate the remaining systematic error due to uncertainty in the value of the (retuned) $\kappa_c$.

\section{Results}
Due to suppression by the heavy quark mass, the form factors $f_+^{D\to P}$ are dominated by the form factors $f_\bot^{D\to P}$, for which our fits are very well-behaved.  By normalizing the form factors to convenient fiducial points, we compare the shapes obtained from LQCD and experiments; this approach eliminates the need for any assumption about the normalization of the experimental data.  Below we overlay fiducially normalized $\chi$PT curves from our fits (S$\chi$PT extrapolated to the physical light quark mass and continuum limit) and form factor shapes from CLEO and BABAR~\cite{CLEO,BABAR}.
\begin{figure}[tbph]
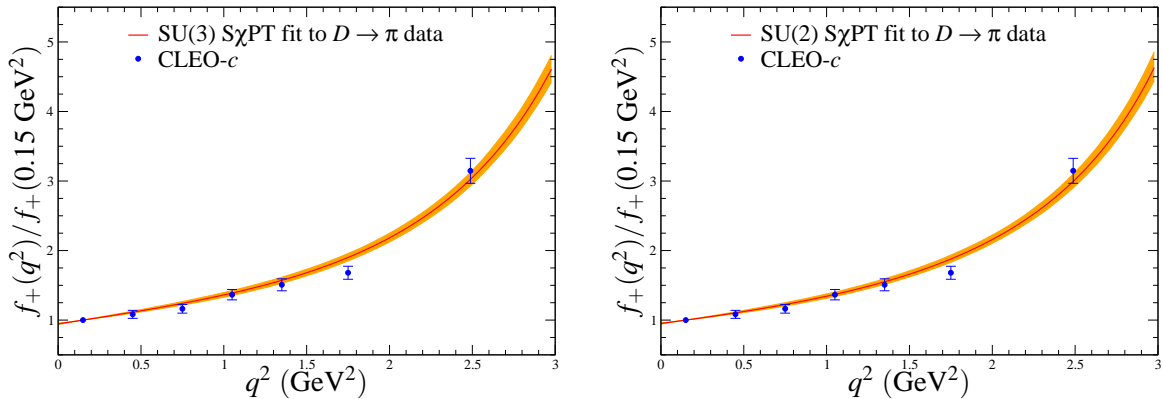

\begin{minipage}[c]{0.47\textwidth}
\begin{center}
\psfrag{fplus_over_fplus_fid}[b][b][1.0]{$f_+(q^2)/f_+(0.15\ \mr{GeV}^2)$}
\psfrag{q2GeV2}[][][1.0]{$q^2\ (\mr{GeV}^2)$}
\includegraphics[clip,width=1.0\textwidth]{shape_check_d2pi_ALL_su3_PRELIM_v2}
\end{center}
\end{minipage}
\hspace{0.05\textwidth}
\begin{minipage}[c]{0.47\textwidth}
\begin{center}
\psfrag{fplus_over_fplus_fid}[b][b][1.0]{$f_+(q^2)/f_+(0.15\ \mr{GeV}^2)$}
\psfrag{q2GeV2}[][][1.0]{$q^2\ (\mr{GeV}^2)$}
\includegraphics[clip,width=1.0\textwidth]{shape_check_d2pi_ALL_su2_PRELIM_v2}
\end{center}
\end{minipage}
\caption{\label{fig:comp_d2pi}Preliminary LQCD results for the shape of $f_+^{D\to\pi}(q^2)$, with statistical errors only, compare favorably with CLEO-$c$ data, with total errors~\cite{CLEO}.  On the left the curve is the chiral-continuum extrapolated shape from SU(3) S$\chi$PT; on the right the curve is from SU(2) S$\chi$PT.  The curves agree within statistics.}
\end{figure}
\begin{figure}[tbph]
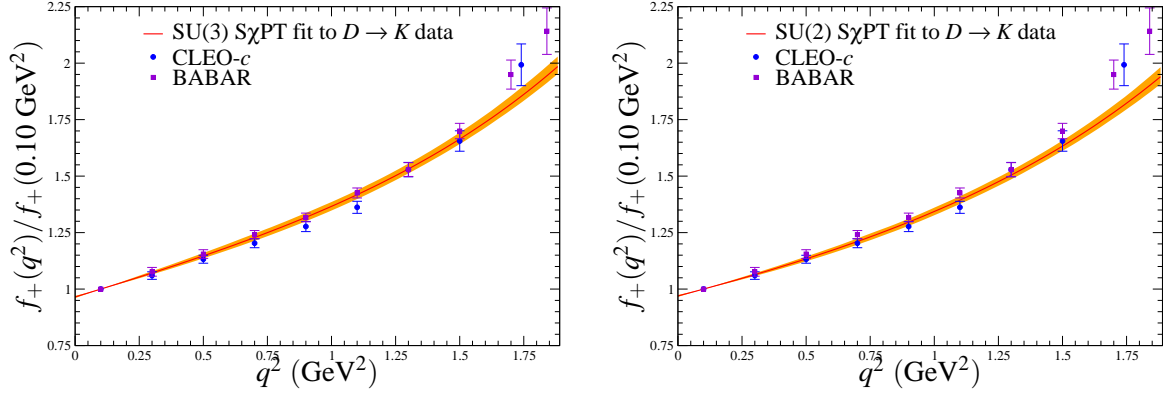

\begin{minipage}[c]{0.47\textwidth}
\begin{center}
\psfrag{fplus_over_fplus_fid}[b][b][1.0]{$f_+(q^2)/f_+(0.10\ \mr{GeV}^2)$}
\psfrag{q2GeV2}[][][1.0]{$q^2\ (\mr{GeV}^2)$}
\includegraphics[clip,width=1.0\textwidth]{shape_check_d2k_ALL_su3_PRELIM_v2}
\end{center}
\end{minipage}
\hspace{0.05\textwidth}
\begin{minipage}[c]{0.47\textwidth}
\begin{center}
\psfrag{fplus_over_fplus_fid}[b][b][1.0]{$f_+(q^2)/f_+(0.10\ \mr{GeV}^2)$}
\psfrag{q2GeV2}[][][1.0]{$q^2\ (\mr{GeV}^2)$}
\includegraphics[clip,width=1.0\textwidth]{shape_check_d2k_ALL_su2_PRELIM_v2}
\end{center}
\end{minipage}
\caption{\label{fig:comp_d2k}Preliminary LQCD results for the shape of $f_+^{D\to K}(q^2)$, with statistical errors only, compare favorably with CLEO-$c$~\cite{CLEO} and BABAR data~\cite{BABAR}, with total errors.  On the left the curve is the chiral-continuum extrapolated shape from SU(3) S$\chi$PT; on the right the curve is from SU(2) S$\chi$PT.  The curves agree within statistics.}
\end{figure}
The errors on the experimental (blue and violet) data points are from the full covariance matrix, including systematics.  Even though we omit systematic uncertainties in the lattice results in the above plots, the qualitative agreement between the curves and experiment is perfectly acceptable for both SU(3) and SU(2) $\chi$PT.  Quantitative tests can be performed by fitting the lattice results and experimental data separately to the $z$-expansion~\cite{BecherHill}.  Once the quantitative compatibility of the lattice and experiment form factor shapes is verified, simultaneously fitting the lattice results and experimental data will yield the CKM matrix elements $|V_{cs}|$ and $|V_{cd}|$.

From the SU(2) (SU(3)) fits above, the statistical errors in $f_+^{D\to\pi}(0)$ are 4\% (4\%), and those in $f_+^{D\to K}(0)$, 2.3\% (3\%).
Important systematic errors are from heavy-quark lattice artifacts, the error in $r_1$, and the error in the axial coupling $g_{\pi}$.  Naively updating the $g_\pi$ error reduces the projected systematics to 3.4\%~\cite{Meinel,lat2011proc}.  A careful estimate of all systematics reflecting the entire data set has yet to be made; the difference between our present SU(2) and SU(3) curves is in some cases comparable to the other errors.

Fermilab is operated under contract DE-AC02-07CH11359 with the U.S. DOE; J.A.B. is supported by the Creative Research Initiatives Program (2012-0000241) of the NRF grant funded by the Korean government (MEST); this work was supported by grants DE-FG02-91ER40661 (S.G., R.Z.) and DE-FG02-91ER40677 (D.D., R.D.J., A.X.K.) of the U.S. DOE.
%

\begin{thebibliography}{99}
\bibitem{PDG} 
  J.~Beringer {\it et al.}  [PDG],
  Phys.\ Rev.\ D {\bf 86}, 010001 (2012).
\bibitem{FNALMILC}
  A.~Bazavov {\it et al.} [FNAL/MILC],
  Phys.\ Rev.\ D {\bf 85}, 114506 (2012)
  arXiv:1112.3051 [hep-lat].
  C.~Aubin {\it et al.} [FNAL/MILC],
  Phys.\ Rev.\ Lett.\  {\bf 94}, 011601 (2005)
  [hep-ph/0408306].
\bibitem{HPQCD}
  H.~Na {\it et al.} [HPQCD],
  Phys.\ Rev.\ D {\bf 86}, 054510 (2012)
  arXiv:1206.4936 [hep-lat].
  H.~Na {\it et al.} [HPQCD],
  Phys.\ Rev.\ D {\bf 84}, 114505 (2011)
  arXiv:1109.1501 [hep-lat].
  H.~Na {\it et al.} [HPQCD],
  Phys.\ Rev.\ D {\bf 82}, 114506 (2010)
  arXiv:1008.4562 [hep-lat].
\bibitem{BABAR}
  B.~Aubert {\it et al.}  [BABAR],
  Phys.\ Rev.\ D {\bf 76}, 052005 (2007)
  arXiv:0704.0020 [hep-ex].
\bibitem{CLEO}
  D.~Besson {\it et al.}  [CLEO],
  Phys.\ Rev.\ D {\bf 80}, 032005 (2009)
  arXiv:0906.2983 [hep-ex].
\bibitem{NPBdecays} 
  R.~Zhou {\it et al.} [FNAL/MILC], these proceedings,
  arXiv:1211.1390 [hep-lat].
  Jon~A.~Bailey {\it et al.} [FNAL/MILC],
  Phys.\ Rev.\ Lett.\  {\bf 109}, 071802 (2012)
  arXiv:1206.4992 [hep-ph].
  S.W.~Qiu {\it et al.} [FNAL/MILC], these proceedings,
  arXiv:1211.2247 [hep-lat].
  C.~M.~Bouchard {\it et al.} [HPQCD], these proceedings,
  arXiv:1210.6992 [hep-lat].
\bibitem{Wingate:2002fh} 
  M.~Wingate {\it et al.} [HPQCD],
  Phys.\ Rev.\ D {\bf 67}, 054505 (2003)
  [hep-lat/0211014].
\bibitem{Aidaetal}
  A.X.~El-Khadra and A.S.~Kronfeld, private communication.
\bibitem{btopipaper}
  Jon A. Bailey {\it et al.} [FNAL/MILC],
  Phys.\ Rev.\ D {\bf 79}, 054507 (2009)
  arXiv:0811.3640 [hep-lat].
\bibitem{lat2009proc}
  Jon A. Bailey {\it et al.} [FNAL/MILC],
  PoS LAT {\bf 2009}, 250 (2009)
  arXiv:0912.0214 [hep-lat].
\bibitem{AB_SHMChPT}
  C.~Aubin and C.~Bernard,
  Phys.\ Rev.\ D {\bf 76}, 014002 (2007)
  arXiv:0704.0795 [hep-lat].
\bibitem{Ran_note_sharpe}
  R.~Zhou, collaboration notes.  See also
  Jon A. Bailey {\it et al.} [SWME],
  Phys.\ Rev.\ D {\bf 85}, 074507 (2012)
  arXiv:1202.1570 [hep-lat].
\bibitem{BecherHill} 
  T.~Becher and R.~J.~Hill,
  Phys.\ Lett.\ B {\bf 633}, 61 (2006)
  [hep-ph/0509090].
\bibitem{Meinel}
  W.~Detmold {\it et al.},
  Phys.\ Rev.\ D {\bf 85}, 114508 (2012)
  arXiv:1203.3378 [hep-lat].
\bibitem{lat2011proc}
  Jon A. Bailey {\it et al.} [FNAL/MILC],
  PoS LATTICE {\bf 2011}, 270 (2011)
  arXiv:1111.5471 [hep-lat].
\end{thebibliography}
\end{document}